\documentclass[pre,aps,nofootinbib,twocolumn,showpacs]{revtex4}
\usepackage{graphics}
\usepackage{epsfig}

\begin{document}
\newcommand{\be}{\begin{equation}}
\newcommand{\ee}{\end{equation}}

\title{One-dimensional spin-anisotropic kinetic Ising model 
subject to quenched disorder}

\author{N\'ora Menyh\'{a}rd$^1$ and G\'eza \'Odor$^2$}
\affiliation{$^1$ Research Institute for Solid State
Physics and Optics, H-1525 Budapest, P.O.Box 49, Hungary \\
$^2$ Research Institute for Technical Physics
and Materials Science, H-1525 Budapest, P.O.Box 49, Hungary}

\begin{abstract}

 Large-scale Monte Carlo  simulations are used to explore the
 effect of quenched disorder on one dimensional, non-equilibrium 
 kinetic Ising models with locally broken spin symmetry, at zero 
 temperature (the symmetry is broken through spin-flip rates that 
 differ for '$+$' and '$-$' spins). The model is found to exhibit a 
 continuous phase transition to an absorbing state. The associated  
 critical behavior is studied at zero branching rate of kinks, 
 through analysis spreading of '$+$' and '$-$' spins and, of the 
 kink density. Impurities exert a strong effect on the critical 
 behavior only for a particular choice of parameters, corresponding 
 to the strongly spin-anisotropic kinetic Ising model introduced
 by Majumdar et al. Typically, disorder effects become evident 
 for impurity strengths such that diffusion is nearly blocked. In 
 this regime, the critical behavior is similar to that arising,
 for example, in the one-dimensional diluted contact process, with 
 Griffiths-like behavior for the kink density. We find variable 
 cluster exponents, which obey a hyperscaling relation, and are
 similar to those reported by Cafiero et al. We also show that the 
 isotropic two-component $AB \to \emptyset $  model is insensitive 
 to reaction-disorder, and that only logarithmic corrections arise,  
 induced by strong disorder in the diffusion rate.

\end{abstract}
\maketitle

 \section{ Introduction }
 The study of non-equilibrium model systems has attracted great attention
 in recent years. A variety of phase transitions have been found
 characterized by critical exponents, both static and dynamic. 
 Of special interest are transitions from a fluctuating active state into 
 an absorbing one.
 A wide range of models with transitions into absorbing states was
 found to belong to the directed percolation (DP) universality
 class \cite{DP,Uclassrev}. 
 Another universality class of interest is the so-called
 parity conserving (PC) class \cite{jen94,dani,kim94,park}.
 The most extensively  studied  model in this class is branching
 annihilating random walks with an even number of offsprings (BARWe)
 in one dimension (1d). The first examples of models exhibiting a 
 PC-type transition were two one-dimensional cellular automata studied 
 by Grassberger \cite{gra8489}. The prototype {\it spin}-model for 
 PC-type phase transitions, the non-equilibrium kinetic Ising model 
 (NEKIM) was proposed by one of the authors \cite{men94}.
 In this model domain walls (kinks) between unlike spins follow
 BARWe dynamics.
 
 One dimensional models showing DP transition, such as the contact 
 process, exhibit pronounced changes in behavior under the effect of 
 quenched spatial disorder, even for small impurity concentrations 
 (see \cite{Vojta} and references there). On the other hand, as a recent 
 computer simulation study \cite{om05} has shown with great accuracy, the 
 parity-conserving class appears to be highly resistant to impurities. The 
 same study reported similar, or even stricter, negative results for the 
 one-dimensional Glauber-Ising model \cite{gla63}.

 Another aspect of the problem, however, was identified by 
 Majumdar et al. \cite{mdg}, who introduced a specific, inherent
 spin anisotropy (kinetic disorder) in the Glauber-Ising model
 (Majumdar Dean Grassberger (MDG) model in the following),
 These authors found, both analytically and numerically,
 a slow logarithmic factor in the decay of the density of kinks 
 $\rho(t)$ for $t\rightarrow \infty$. 

The aim of the present study is to show
that in certain Ising-like systems (possessing two absorbing states),
local kinetic disorder may effectively remove one of the absorbing states.
The resulting, single-absorbing state system is sensitive
to spatial disorder, similar to models in the directed percolation class.
To this end we start from a generalization of NEKIM,
the nonequilibrium kinetic Ising model with anisotropy (NEKIMA)
\cite{mo02,mo03}, with variable kinetic disorder, and
add uncorrelated spatial impurities to the system.
We recall that without spatial impurities and in the zero-branching limit
considered throughout this paper, the exponent $\alpha$, which governs the
time dependence of the kink density, $\rho(t)\sim t^{-\alpha}$, is
invariably $\alpha =1/2$ (apart from possible logarithmic corrections).
The characteristic cluster-exponents, which are more susceptible to
kinetic disorder than those governing the kink density, differ however
from their Glauber-Ising values. This phase will be called
spin-anisotropic Glauber-Ising (SAGI). For definitions and
details of the models referred to here, and the acronyms used,
see Table \ref{models}.
\begin{table*} \label{models}
\begin{ruledtabular}
\caption{Summary of models. Here $w$ denotes the spin transition
  probabilities, $A=\Gamma(1+\tilde\delta)$, $B=\Gamma/2(1 - \tilde\delta)$}
\begin{tabular}{|c|c|c|c|c|c|c|} 
Transition & $w(+,--)$ & $w(-,++)$  &  $w(-,+-)=$    & $w(+,+-)$& $w_{ex}(
s_i,s_{i+1})=$ & kink \\ 
probability &           &            &  $w(-,-+)=p$ & $w(+,-+)=p_+$ & $p_{e
x}(1-s_is_{i+1})$ & dynamics \\ \hline
Glabuer- & $A$ & $A$ & $B$ & $B$ & $0$ & ARW \\
Ising    & $\tilde\delta\ge 0$ & $\tilde\delta\ge 0$ &      &  & & \\  \hline
NEKIM   & $A$ & $A$ & $B$ & $B$ & $\ne 0$ & PC  \\
        & $\tilde\delta < 0$ & $\tilde\delta < 0$  & & &    & \\ \hline
NEKIMA  & $A$ & $0$ & $p$ & $p_+=p=1/2$ (MDG) &      $\ne 0$ & DP  \\
        & $\tilde\delta=0, \Gamma=1$ & (MDG) & $(=1/2)$ & $p_+ < p$ (SAGI) & $0$ & SAGI \\ \hline
NEKIMCA  & $A$ & $A$ & $B$ & $B$ & $0$  & PC \\
CA update &    &     &     &     &           & \\
\end{tabular}
\end{ruledtabular}
\end{table*} 

Our numerical studies show that the scaling behavior of the SAGI phase is
resistant to quenched spatial impurities, of
strength $p_0$, in the small-impurity regime.
Close to the incipient freezing of a diffusion
channel, which occurs at about $p_0=0.5$, and which appears to be a dirty
critical point \cite{Vojta}, the exponent $\alpha$ jumps from $0.5$ to
$0.21$. On increasing $p_0$ further,
a Griffiths-like \cite{Griffiths} phase is encountered, with $\alpha(p_0)$
increasing with $p_0$. The cluster properties: the mean population size
$n(t)\sim t^\eta$, the mean-square distance of spreading
of spins $R^{2}(t) \sim t^z$, and  the survival probability,
$P(t)\sim t^{-\delta}$ are also found to change from their SAGI
expressions (Table \ref{clusres}) to those characteristic of a Griffiths phase
\cite{Griffiths}. We find close similarities to the results on
scaling in the impure contact process \cite{Cafi} in (almost) the
whole plane of the corresponding phase diagram, with
the hyperscaling law remaining valid.
Here this behavior is found in an annihilating random walk-type
model (ARW) i.e.,  without kink production, while production takes place
on the level of spins, in one of the spin-channels
('$+$' spins in the following), as explained in Sect. II.

Moreover, in the present case  the  SAGI phase plays the role of a
supercritical phase: active-like  concerning '$+$'- cluster-behavior
while frozen for the '$-$' spin phase, and of annihilating-random-walk 
type  for the kink density $\rho(t)$.
The dual nature of global and cluster properties is not 
contradictory and can be traced back to the fact that the 
model is  highly spin-anisotropic.

\begin{center}
\begin{figure}[ht]
\epsfxsize=80mm
\epsffile{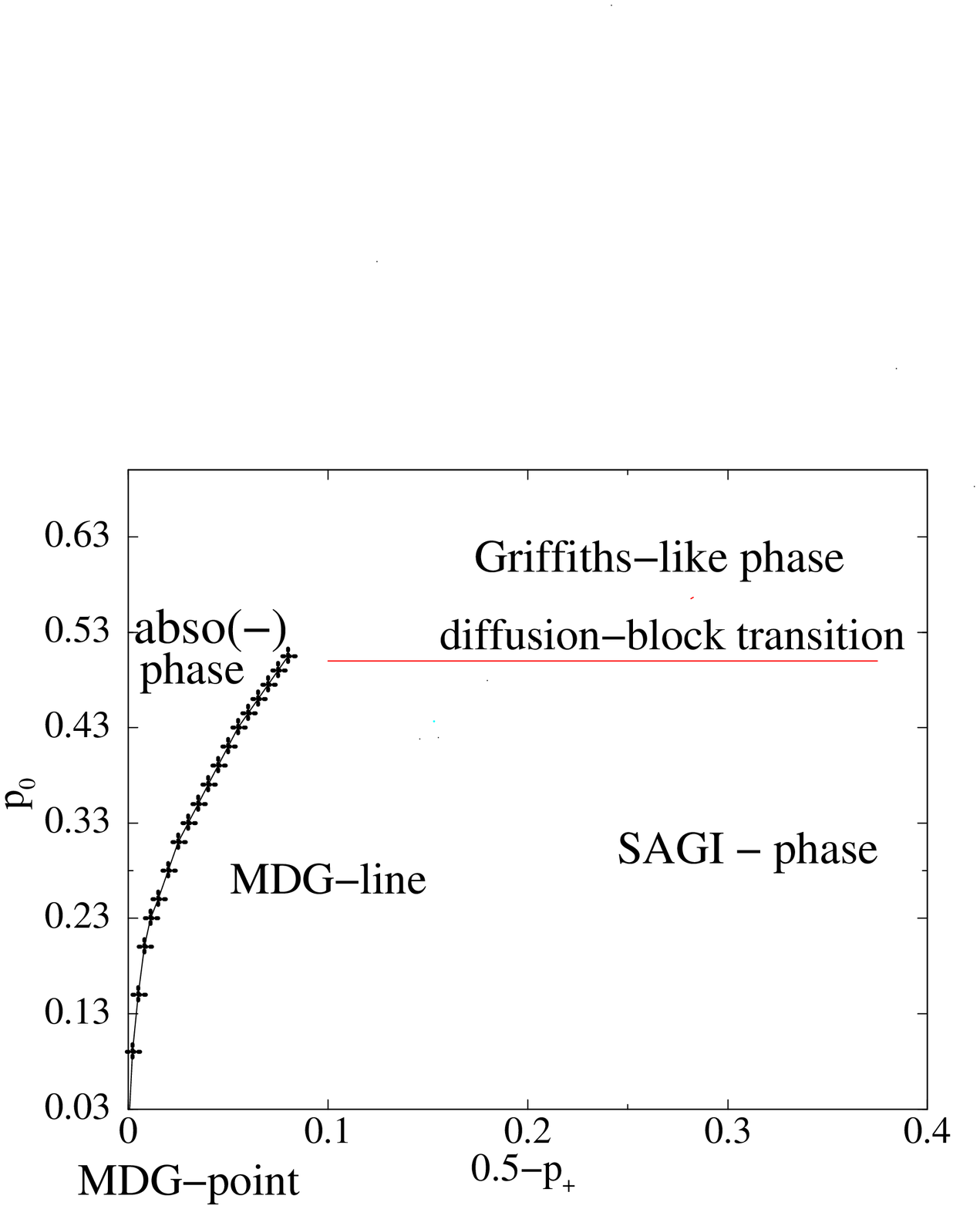}
\vspace{4mm}
\caption{Phase diagram of the impure NEKIMA model ($p_{ex}=0$, ARW case)
Different regions of behavior are marked (for details see text) The
vicinity of $p_+=0$ is not shown as not discussed here.}
\label{phasedia}
\end{figure}
\end{center}

The NEKIMA model coincides, at some specific parameter values (MDG case, 
Fig.\ref{phasedia}) with the model of Majumdar et al. \cite{mdg} 
mentioned above. 
The joint effect of kinetic and spatial impurities is even more
pronounced at and in the vicinity of this point (see the phase
diagram depicted in Fig.\ref{sagipicture}, where the most spin-anisotropic 
point, the MDG-point is at the origin).
In this region of the $(p_{+},p_0)$ plane the impure scaling behavior 
is different: by increasing $p_0$ we
have found a change in the behavior of the global quantity $\rho(t)$: 
 $\alpha $ switches from a value of $1/2$ (apart from the very late time 
logarithmic correction) to $\alpha=1$. The cluster
behavior is different for the '$+$' and '$-$' spins. For the '$+$' it
appears to follow the above mentioned scheme \cite{Cafi} of the impure 
contact process: upon increasing $p_0$ a strong tendency towards the 
absorbing '$+$' phase can be seen with  the hyperscaling law 
\cite{DiTre} satisfied. For the '$-$' cluster we find $\eta=z/2=1$, 
and $\delta=0$, with logarithmic corrections in time.
 
In summary, we  find that  by increasing spin anisotropy the 
sensitivity of critical behavior to quenched disorder is enhanced. 
(NEKIM $\to$ SAGI $\to$ MDG). It is worth noting that in order to 
maintain the symmetry of the NEKIMA model, the disorder affecting 
diffusion is anisotropic too, therefore it is different from the 
problem considered in \cite{om05}, where for very strong disorder 
diffusion can be blocked completely..

In the Appendix a further, isotropic ARW model,
$AB \to\emptyset$, \cite{DP} is investigated via numerical simulations,
under quenched disorder. In this model, which can be
mapped onto the pure MDG model \cite{Uclassrev} the effect of impurities
on the annihilation and diffusion rates are investigated separately: the
density decay is unaffected in case of the former, while
logarithmic corrections arise for impure diffusion rates for very long
times.

The paper is organized as follows. In Sect. II. we define the models and
the quantities to be investigated. Some previous results 
for the exponents of spreading from a localized source are recalled and 
commented on. In a subsection -- to enlighten the different roles played
by the '$+$' and '$-$' spins -- our model is mapped onto a
reaction-diffusion system. In Sect.III. the effect of spatial impurities 
is investigated for two different choices of the value of the 
anisotropy-parameter: the SAGI case (A) and the more spin-anisotropic 
MDG case (B).  The results are discussed in Sect.IV., while
in the Appendix a further ARW model, the $AB\to\emptyset$ is 
investigated under the influence of quenched disorder.

\section{Spin-anisotropic Glauber-Ising model.}

In the one-dimensional Glauber  model at temperature zero, the most general 
form for the flipping rate of spin $s_i$ is  \cite{gla63} 
($s_i=\pm1$):
\begin{equation}
w(s_{i},s_{i-1},s_{i+1}) ={\Gamma\over{2}}(1+\tilde\delta s_{i-1}s_{i+1})
[1 - {1\over2}s_i(s_{i-1} + s_{i+1})]
\label{Gla}
\end{equation}
Usually the Glauber model is understood as the special case  
$\tilde\delta=0$, $\Gamma=1$
and we will use these parameter values in the following.
Processes involving the reaction 1 kink $\rightarrow$ 3 kinks 
are introduced via the exchange rate
\begin{equation}
w_{ex}(s_i,s_{i+1})={ p_{ex}\over{2}}(1-s_{i}s_{i+1})
\label{Kaw}
\end{equation}
(a kink is a '$-+$', or '$+-$' configuration: domain boundary between
two oppositely magnetized regions).
For negative values of $\tilde\delta$ this model (called NEKIM), 
shows a line of PC-transitions in the  ($p_{ex}$, $\tilde\delta$)-plane 
\cite{men94}. The NEKIMA model is an extension with local symmetry 
breaking in the flipping rates of the  '$+$' and '$-$' spins, as follows. 
Concerning the annihilation rates
(of a spin in the neighborhood of oppositely oriented spins)
the prescription in \cite{mdg} is followed:
\begin{equation} 
\label{mdgrat}
w(+;--)=1, \,\,  
w(-;++)=0. 
\end{equation}
Further spin symmetry breaking is introduced
in the diffusion part of the Glauber transition rate as follows.
The transition rates
\begin{equation}
p\equiv w(-;+-)=w(-;-+)=\Gamma /2 (1-\tilde\delta)
\label{orig}
\end{equation}
are taken as in  Eq.(\ref{Gla}), while
$w(+;+-)$ and  $w(+;-+)$ may  take smaller values:
\begin{equation}
p_+\equiv w(+;+-)=w(+;-+)\leq p.
\label{p+}
\end{equation}
In this way, by locally favoring  $+$ spins, the effect of the 
other dynamically-induced field arising from the rates
(Eq.\ref{mdgrat}) is counterbalanced.
The spin-exchange part of the NEKIM model remains unchanged, Eq.(\ref{Kaw}). 

Spreading from a localized source at criticality is usually described 
by the following three quantities
\begin{eqnarray} \label{clex}
P(t)     \sim t^{-\delta},\nonumber \\ 
{n(t)}    \sim t^{\eta}, \nonumber \\
{R^2(t)}  \sim t^z \ ,        
\end{eqnarray}
where $n(t)$ denotes the mean population size, $R^2(t)$ is the
mean square spreading of particles (here spins) about the origin and
$P(t)$ is the survival probability.
In most cases these quantities are defined for particles, 
in the present case, however, they will be used for spins.

In NEKIMA '$+$' and '$-$' spins are not symmetric, therefore we
have investigated two kinds of clusters. Namely, the development
of the '$-$' cluster-seed was started from a wholly '$+$' environment
while that of the  '$+$' cluster from a sea of '$-$' spins.
We will call them '$-$' cluster and '$+$' cluster, respectively.
\begin{table*} [ht] \label{clusres}
\begin{ruledtabular}
\caption{Cluster critical exponents at the SAGI transition
point and at the MDG point.}
\begin{tabular}{|c|c|c|c|c|c|c|} 
 exponents & SAGI'$+$' & SAGI'$-$' & MDG'$+$' & MDG'$-$'&
 diso MDG'$+$' & diso MDG'$-$'\\ \hline
$\eta$ &1 & 0 & 0 & 0.5 & -1 & 1 \\
$\delta $  & 0 & 0  & 0.5 & 0 & 1 & 0 \\
$ z  $   & 2 & 0 & 1 & 1 & 0 & 2 \\ 
\end{tabular}
\end{ruledtabular}
\end{table*}
The cluster critical exponents  for the NEKIMA were reported in \cite{mo03}.
Their values at a typical SAGI  point  ($p_{+} =0.3, p=0.5$)
 and at the MDG point ($p_{+} =p= 0.5$),
for  the ARW-case  discussed here, are summarized on Table \ref{clusres}.
It should be noted here, however, that the precision of the 
simulations in the above cited paper does not exclude the possibility 
of the presence of some  $\log(t)$ corrections.

On the level of '$+$' - spins the SAGI phase ($1/2=p>p_+$) is an active one
as shown by the '$+$'-cluster exponents on Table \ref{clusres}. 
For producing '$+$' spins the rate in Eq.(\ref{orig}) with $p=1/2$ is 
responsible, while $p_+$ annihilates them. Crudely speaking the 
probability of the '$+$' spin-generating transition process, 
$p-p_{+}$, can be thought of as the 
quantity corresponding to the particle production rate in the contact process. 

As one can see due to the anisotropies the '$-$' and '$+$' spins follow 
different dynamics. To understand the peculiar behavior of the 
species we transform their motion into the language of
reaction-diffusion models, making a connection with the basic models 
and classes \cite{Uclassrev}.

\subsection{Particle picture of the spin-processes}

Let X be a particle and $\emptyset$ its absence. 
For the '$-$' spins as X-es and + spins as $\emptyset$-s 
Eq.(\ref{mdgrat}) gives
\begin{equation}
X\emptyset X \to XXX
\label{s1}
\end{equation}
creation, while Eq.(\ref{p+}) leads to the creation
\begin{equation}
X\emptyset\emptyset \to XX\emptyset, \ \ \emptyset\emptyset X\to\emptyset XX
\label{s2}
\end{equation}
From the above three, the net process remains as
\begin{equation}
\emptyset X \to XX
\label{s3}
\end{equation}
Similarly Eq.(\ref{orig}) means the coagulation processes
\begin{equation}
\emptyset XX \to \emptyset\emptyset X, \ \
XX\emptyset \to X\emptyset\emptyset \ .
\label{s4}
\end{equation}
From eqs.(\ref{s3}) and (\ref{s4}) the generic reactions
\begin{equation} 
X \leftrightarrow 2X
\label{reversiblemod}
\end{equation}
follow. This corresponds to the recently studied 
reversible model, \cite{bur,kamen,jms06,hbm06}, where transition to an 
absorbing phase can occur only for zero branching rate, which is 
excluded in our case, hence we don't expect an active-absorbing 
phase transition of the '$-$' spins.

On the other hand for the '$+$' spins as X's we get from
Eq.(\ref{mdgrat}) the
\begin{equation}
\emptyset X\emptyset \to \emptyset\emptyset\emptyset
\label{annih} 
\end{equation}
spontaneous decay and from Eq.(\ref{p+}) the coagulation
\begin{equation}
\emptyset XX \to \emptyset\emptyset X, \ \
XX\emptyset \to X\emptyset\emptyset \ ,
\end{equation}
while Eq.({\ref{orig}) gives the creation by a neighbor
\begin{equation}
X\emptyset\emptyset \to XX\emptyset, \ \ \emptyset\emptyset X\to\emptyset XX
\end{equation}
These give again $XX \leftrightarrow X$ but according to Eq.(\ref{annih}) 
here the spontaneous annihilation of X, $X\to\emptyset$ is also possible
due to Eq.(\ref{mdgrat}), which is a necessary condition for a DP class
transition. Hence for '$+$' spins a phase transition (at finite reaction
rates) is not excluded and as we will see later it emerges as the
effect of the disorder. The corresponding critical behavior is
similar to that of the disordered DP (albeit an anisotropic one).

\section{Effects of quenched impurities}

At each site $i$ the diffusion rates of kinks $p=w(-;+-)=w(-;-+)$
in Eq.(\ref{orig}) are modified by adding quenched, uncorrelated impurities 
with uniform probability distribution of the form
\begin{equation}
p_i = p + p_0*(2\epsilon(i)-1) \ ,
\label{p0}
\end{equation}
where $\epsilon(i)$ is randomly distributed in the interval 
$0\leq \epsilon(i)\leq 1$ and $0\leq p_{0}\leq 0.5$.
On the other hand the rates, which fix the (a)symmetry of the model 
$w(+;--)=1$, $ w(-;++)=0$ and $p_+$ of Eq.(\ref{p0}) are kept uninfluenced.
In this way kinetic impurities and quenched local impurities
are let to act separately, though simultaneously, resulting in a
rich phase diagram. 

Two characteristic impure behaviors are  exemplified by the following
two choices of the fixed parameters: 
\noindent(1) $p_+=0.3$ (to be called the SAGI-case) and 
\noindent(2) $p_+=0.5$ (to be called the MDG-case).

The phase diagram of the impure NEKIMA is shown schematically on
Fig.\ref{phasedia}. At the MDG point the '$-$' phase gets 
dominating (Section III.B.). 
This absorbing '$-$' - phase is separated from the SAGI phase
by the MDG-line, a line of critical points characterized by the
cluster exponents given in Table \ref{clusres} and where
$\alpha =1/2$. 
This line is depicted here only tentatively (especially for higher
$p_0$ values), as its neighborhood is not the subject of our further 
detailed study. The diffusion-block transition line corresponds 
to the limiting value: $p_0=1/2$, where the only non-fixed 
spin-flip probability can become zero, hence diffusion blocking starts. 
The SAGI phase in the vicinity of this point will be discussed in the
next Section. 
It is worth emphasizing again that in the present model only one 
transition probability gets impure (one spin-flip rate) in contrast to  
a previous investigation of NEKIMCA with impurities \cite{om05}, 
where all four  transition rates 
(spin-flip+annihilation) have been complemented with a term like that
in Eq.(\ref{p0}). We have also looked into the effect of impurities in one of
the annihilation rates of the two fixing the model's anisotropy: no
effect shows up in the vicinity of $p_0=0.5$ in this case as can be expected.

\subsection{The SAGI case ($p_{+}=0.3$)}  

We will investigate the effect of impurities below the MDG point 
i.e. for $p_{+} < 0.5$ at a typical parameter value $p_+=0.3$ 
(see Fig.\ref{phasedia}) by simulating a system of size 
$L=3\times 10^4$ with periodic boundary conditions up to 
$t_{max}=10^6$ Monte Carlo steps (MCS) 
(throughout the paper time is measured by MCS).
For small impurity rates $p_0$ the behavior is found unperturbed. 
Correspondingly, the critical exponent of the 
density of kinks as a function of time is $\alpha=1/2$ and the
exponents of the $'$+$'$ and $'$-$'$ clusters are as seen on 
Table \ref{clusres}. 
By increasing $p_0$, the rate of disorder, however, the spin flip rate 
$p_i$ in Eq.(\ref{p0}) can become zero in the vicinity of $p_0=0.5$ 
and above. To interpret the results in this region it is important 
to recall \cite{mo03} that the critical behavior of the kink density 
and that of the '$+$' and '$-$' clusters are not connected
in the present model (thus no scaling law connects the critical
dynamical exponent $Z$ and the cluster size exponent $z$).
While there is no particle production on the level
of kinks (ARW-process), on the level of spins there is production as 
mentioned in the previous section. The spin-transition rate $p$,
for the process $(- - +) \to (- + +) $ leads to the  increase of $+$
spins. This $+$-spin production is maximal at $p_0=0$. 
The SAGI phase is an active one, more precisely a '$+$'-active one
(the '$-$'-spin phase is frozen). At the same time in the usual sense 
of Glauber-Ising-type models, on the level of kinks, there is no 
active phase  as the kink production $p_{ex}$ in Eq.(\ref{Kaw})
is zero.

By increasing $p_0$ the system stays in this '$+$'-active phase,
neither the kink-decay exponent $\alpha=1/2$ nor the cluster
exponents of Table \ref{clusres} change. Approaching $p_0=1/2$, however, 
the impure spin-flip term  $p_i$, Eq.(\ref{p0}),
can become zero at certain sites and the process diagram
becomes as shown on Fig. \ref{sagipicture}.
\begin{figure}[ht]
\epsfxsize=80mm
\epsffile{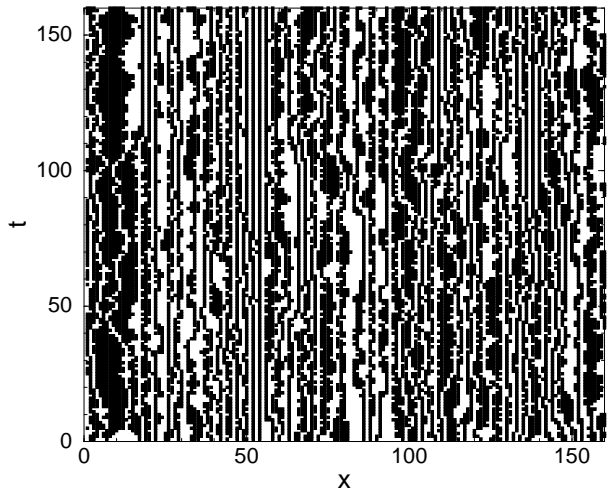}
\caption{Space-time picture at $p_+=0.3$, $p_0=0.5$ (disordered SAGI)
starting from a random initial distribution of '$+$' and '$-$' spins. 
The '$+$'-es are black}
\label{sagipicture}
\end{figure}
The presence of active zones between otherwise frozen-in strips is 
clearly seen there. The similarity to the time-evolution pictures 
(Fig.3 in \cite{Webman}) showing growth of DP clusters in the impure 
active and glassy phases is evident.
These can be thought of as the much-cited active zones of the 
impure contact process \cite{noest,Cafi,Webman,Vojta} inside of the 
otherwise inactive substrate. Thus we can expect similar impure 
critical behavior to take place. This means the generic presence of 
scale invariance, and the existence of a sub-linear growth regime.
This regime is entered now from the active one by increasing $p_0$,
through the 'dirty' critical point at $p_0=0.5$ (no clean critical point exists
in our case, of course).
\begin{figure}[ht]
\epsfxsize=80mm
\epsffile{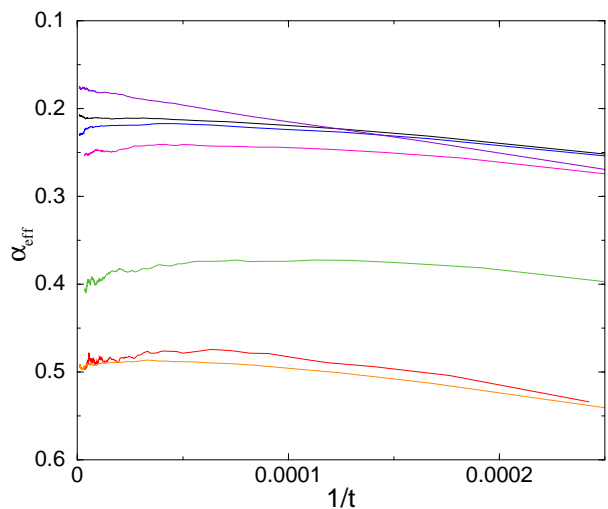}
\caption{Local exponent values of the kink density in the disordered
SAGI phase for different values of the impurity strength 
$p_0 = 0.56, 0.501, 0.503, 0.499, 0.47, 0.4$ and $0.1$ (top to bottom).}
\label{ro3}
\end{figure}
In the vicinity of $p_0=0.5$ the local exponent values defined as
\begin{equation}
\alpha_{eff}(t) = {- \ln \left[ \rho(t) / \rho(t/m) \right] 
\over \ln(m)} \label{slopes}
\end{equation}
(where we used $m=4$) behave as on Fig.\ref{ro3}. 
The fixed point value of the density of kinks at 
$p_0=0.5$ can be fitted with a power of $\ln(t)$
\begin{equation}
\rho(t)\sim{(lnt)}^{-\tilde\alpha} \ ,
\label{dirtyscaling}
\end{equation}
with $\tilde\alpha=1.45943$ as shown on Fig. \ref{rolnfit}. This
exponent differs from that of the known value of the infinite
randomness fixed point of the 1+1d DP ($\tilde\alpha_{DP}=0.38197$) 
\cite{hoy}, not surprisingly kinks follow an anisotropic 
dynamics here.
\begin{figure}
\epsfxsize=80mm
\epsffile{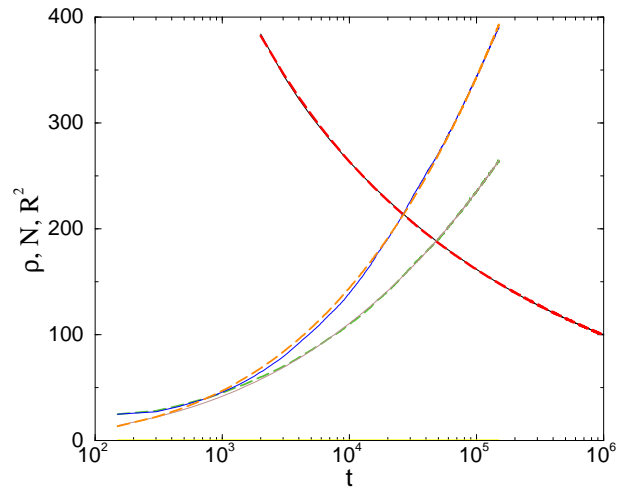}
\caption{The "dirty critical point" of the intermediate SAGI phase
entered by increasing $p_0$ at $p_0=0.5$ is demonstrated by this
fit. The decreasing curve corresponds to the kink density; the higher
increasing curve shows the spreading radius $R^2$, while the lower 
increasing one the spin number $N$ of '$+$' spin domains. The dashed
lines on top the solid ones show activated scaling fit of the form
(\ref{dirtyscaling}).} 
\label{rolnfit}
\end{figure}
The excellent fit with Eq. (\ref{dirtyscaling}) suggests that 
this point is a dirty critical point, which is reached from the 
'active' phase (here from the '$+$'-active phase). 
The value of $\alpha$ is a new critical exponent.
For higher values of impurity strength a Griffiths-like phase with
\begin{equation}
\rho(t)\sim t^{-\alpha(p_0)}
\end{equation}
is entered. Though in  this region the present formulation of the 
impure problem already goes beyond its range of validity, however
the  tendency of the impure behavior is well exhibited in our simulations
(e.g. $\alpha(p_0)$ increases steadily as expected in a 
Griffiths phase, etc.). Nevertheless, we refrain from  giving
more details.

The time dependences of the $'$+$'$ and $'$-$'$ cluster quantities 
(Eqs. \ref{clex}) also show a behavior characteristic 
of an impure phase transition.
The behavior of the $'$-$'$ cluster shows an effect of 'melting':
$\eta$ and $z/2$ move from $0$ to some finite values,
upon increasing $p_0$, while $\delta=0$ remains the same.
The $'$+$'$ cluster behavior however, shows a scenario of a
transition to an absorbing state, similarly to the contact 
process investigated by Cafiero et al. \cite{Cafi}. 
However the direction is opposite: in \cite{Cafi} for small 
values of disorder (called $p$ in their paper) 
an absorbing phase is present, while the active is reached at $p=1$. 

We plot on Fig. \ref{deltauj} the '$+$' cluster behavior, namely 
the local exponent values of the survival probability defined as
\begin{equation}
\delta_{eff}(t) = {- \ln \left[ P(t) / P(t/m) \right] 
\over \ln(m)} \label{slopesd}
\end{equation}
for different values of disorder.
\begin{figure}
\epsfxsize=80mm
\epsffile{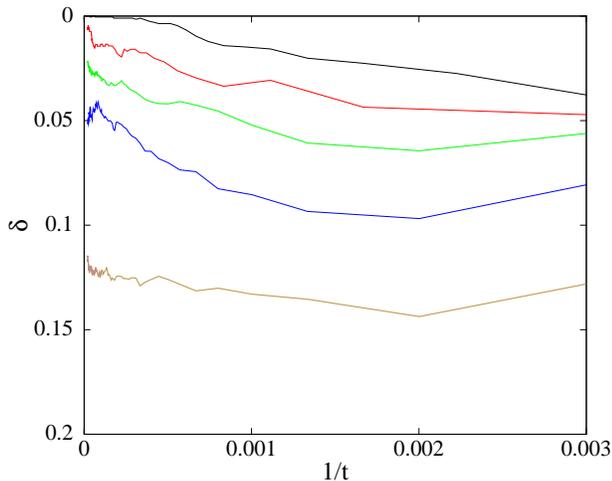}
\caption{Local exponent values of $P(t)$ of the '$+$' clusters
in the disordered SAGI phase for: 
$p_0 = 0.47, 0.5, 0.51, 0.525, 0.56$ (from top to bottom).}
\label{deltauj}
\end{figure}
One can clearly see a turn from the SAGI-active behavior ($\delta=0$) 
towards a Griffiths-phase-like non-universal cluster scaling 
by increasing $\delta$ in the vicinity of $p_0= 0.5$.

At $p_0= 0.5$ (dirty critical point of the '$+$' spins) 
an ultra-slow, activated scaling of the form Eq.(\ref{dirtyscaling}) 
can be observed in the cluster data (see Fig. \ref{rolnfit}). 
Fitting with the activated scaling forms
\begin{equation}
\delta(t) \sim{(lnt)}^{-\tilde\delta} \ ,
R^2(t) \sim{(lnt)}^{\tilde z} \ ,
N(t) \sim{(lnt)}^{\tilde\eta} \ ,
\label{dirtyRN}
\end{equation}
for the '$+$' data resulted in: $\tilde\delta=0.060(6)$, 
$\tilde z = 3.88(1)$ and $\tilde\eta = 3.37(2)$. 
These values again differ from those of the strong disorder fixed point 
of DP, probably due to the anisotropies of the processes.

Our simulation data for the mean size of the '$+$'-cluster 
$R(t)$ depicted on a double-logarithmic scale is shown on Fig.\ref{zuj}. 
The linear SAGI behavior crosses over to decreasing values of $z/2$ 
by reaching the $p_0=0.5$ transition point 
(sub-linear behavior, for similar behavior see Tables I. and 
II. in  \cite{Cafi})
\begin{figure}[ht]
\epsfxsize=80mm
\epsffile{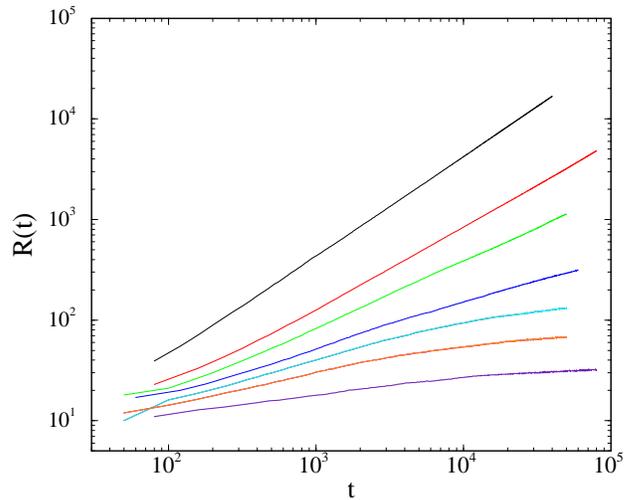}
\caption{$\ln R(t)$ versus $\ln(t)$ for the '$+$' -cluster
in the SAGI phase for: $p_0=0.3,0.45,0.48,0.5,0.52,0.55,.056$ 
(top to bottom)}
\label{zuj}
\end{figure}

\subsection{The MDG-case ($p_+=0.5$)}

The MDG point can be found at the origin of the phase diagram 
(Fig.\ref{phasedia}). At these parameter values the impure system 
decays into a fully compact state, where all spins are flipped to 
'$-$' but different types of kinks exert marginal perturbation 
on each other \cite{om2000}. 
In more detail \cite{mdg} found analytically 
that the magnetization decays as
\begin{equation} 
m(t)= -1 + {\rm const}/\ln(bt) \ ,
\end{equation}
while the kink density behaves as
\begin{equation}
\rho(t) \sim t^{-1/2} \ln(bt)
\end{equation} 
(where $b$ is constant). 
For the time being -- since much stronger 
(power-law-type) effects enter due to the quenched spatial impurities 
-- we disregard the $\ln(t)$-type of correction and show only the
leading singular behavior of the cluster properties 
(see Table \ref{clusres}).

By introducing impurities of strength $p_0=0.3$, the space-time
evolution of kinks looks as can be seen on Fig.\ref{mdgrajz}. 
The absorbing nature of the '$+$' phase is apparent.
\begin{figure}[ht]
\epsfxsize=80mm
\epsffile{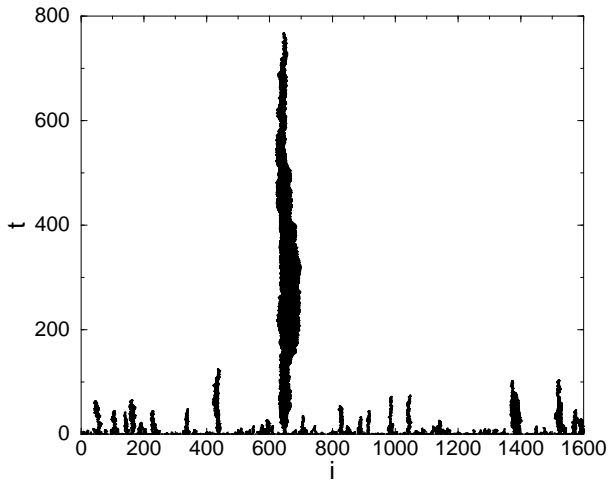}
\caption{Space-time evolution in the disordered MDG model
($p_+=0.5$, $p_0=0.3$) from the random initial configuration. The
$+$ (black) and $-$ (white) spin clusters are separated by kinks}
\label{mdgrajz}
\end{figure}
The fully '$-$' spin configuration sets in quickly (compare this 
figure with Fig. 1. of \cite{mo03}, where single '$-$' spins 
move via random walk in the see of '$+$'-es.).

Results of high-precision simulations are shown on Fig.\ref{mdgsl}.
For $p_0 < 0.5$ one can see $\alpha \to 1$ in the long time limit.
\begin{figure}[ht]
\epsfxsize=80mm
\epsffile{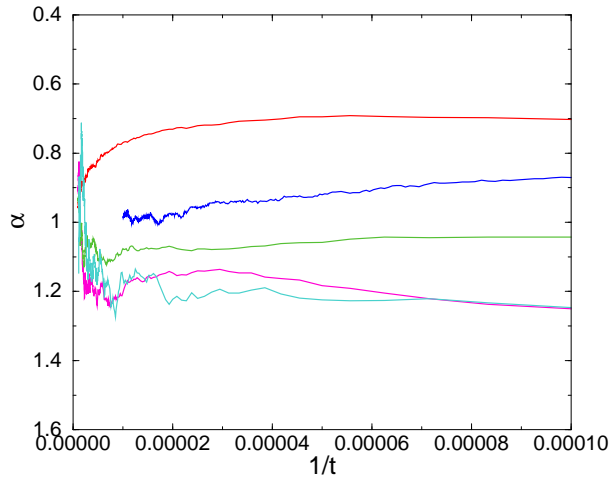}
\caption{Local exponent values of $\rho(t)$ for different values 
of $p_0 = 0.08, 0.18, 0.25, 0.45, 0.5$ (from top to bottom)
indicating $\alpha = 1$--type power-law behavior.}
\label{mdgsl}
\end{figure}
This behavior can be ascribed to the choice of the impurity, which acts 
against one type of diffusion (which favors '$+$' spins). As a consequence 
the '$-$'-es can spread better than in the impure case.
The effect of impurities on the '$-$' clusters is illustrated on 
Fig. \ref{mdgnegcl} for the $\eta$ and $z/2$ exponents. The
local slope curves of $\eta$ and $z/2$ are almost the same,
they approach the asymptotic value $\eta=z/2=1$ from below with 
an upward curvature ($\delta=0$ remains constant).
This kind of effective exponent behavior usually corresponds to 
logarithmic correction to scaling. In case of the general 
form for $N(t)$ 
\begin{equation}
N(t) = t^{\eta} \ln^{b}(t) \ , 
\end{equation}
the effective exponent behaves as
\begin{equation} \label{logform}
\eta_{eff} = \frac {d \ln(N(t) ) } {d \ln(t) } = 
\eta + \frac {b} { \ln (t) }  \ . 
\end{equation}
Therefore a logarithmic fit of the form
\begin{equation}
a + b/\ln(t)
\label{negcluslogfit}
\end{equation}
has been applied for the local exponent curves and the values obtained
both for $\eta$ and $z/2$ are shown in Table \ref{logfitt}. 
\begin{table}[ht]
\caption{Logarithmic fitting results for the '$-$' spins clusters
at the MDG point for various disorder via Eq. (\ref{negcluslogfit}).} 
\vspace{1mm}
\begin{tabular}{|c|c|c|c|}
\hline
     & $p_0=1$ & $p_0=0.56$ & $p_0=0.35$ \\ \hline
$\eta=z/2$ &1.001(1) & 1.004(5) & 0.95(6) \\
$b$   & -1.511  & -1.825 & -1.87\\
\hline
\end{tabular}
\label{logfitt}
\end{table}
The clusters are compact, of course, thus the hyperscaling law 
is satisfied. 

To summarize, instead of the weak logarithmic tendency towards the 
dominance of the '$-$'-es as for the pure MDG model the impurity pushes 
the system into a more anisotropic decay, characterized by the 
kink decay exponent $\alpha = 1$. 
The depletion of the '$+$'-cluster is illustrated on 
Figs.\ref{mdgplclz} and \ref{mdgplcld}.
The '$+$'-cluster dies out apparently in a power-law manner with
the exponents $\eta= -1$, $\delta=1$ and $z/2=0$. 
These scaling behaviors suggest a transition 
from one absorbing phase (MDG) to an other one 
as the effect of quenched impurities, which increases anisotropy.

It is worth noting that by switching on the spin-exchange term ($p_{ex}$)
in the pure case a quicker dominance of the '$-$'-es takes place for
the same parameter values of the model as here.
In that case, no power-law regime is visible (except for some early time 
transient region and for very low values of the kink production 
$p_{ex}$) and the absorbing phase sets in exponentially fast typically 
(see \cite{mo02}) due to the faster destruction of
metastable '$+$' domains.
\begin{figure}
\epsfxsize=80mm
\epsffile{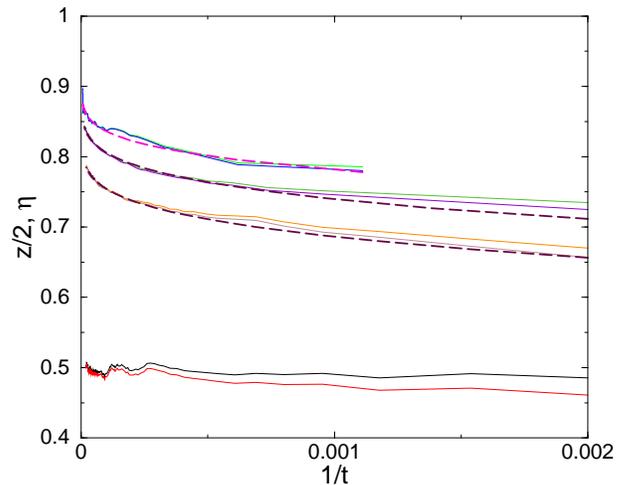}
\vspace{1mm}
\caption{Negative cluster exponents $\eta$ and $z/2 $ in the impure MDG
case for $p_0 = 1, 0.56, 0.35, 0 $ (in pairs, from top to bottom). 
$\delta=0$ ($P(\infty)=1$)  in the whole interval. The dashed lines are
logarithmic fitting of the form (\ref{negcluslogfit}).} 
\label{mdgnegcl}
\end{figure}
\\
\\
\begin{figure}
\epsfxsize=80mm
\epsffile{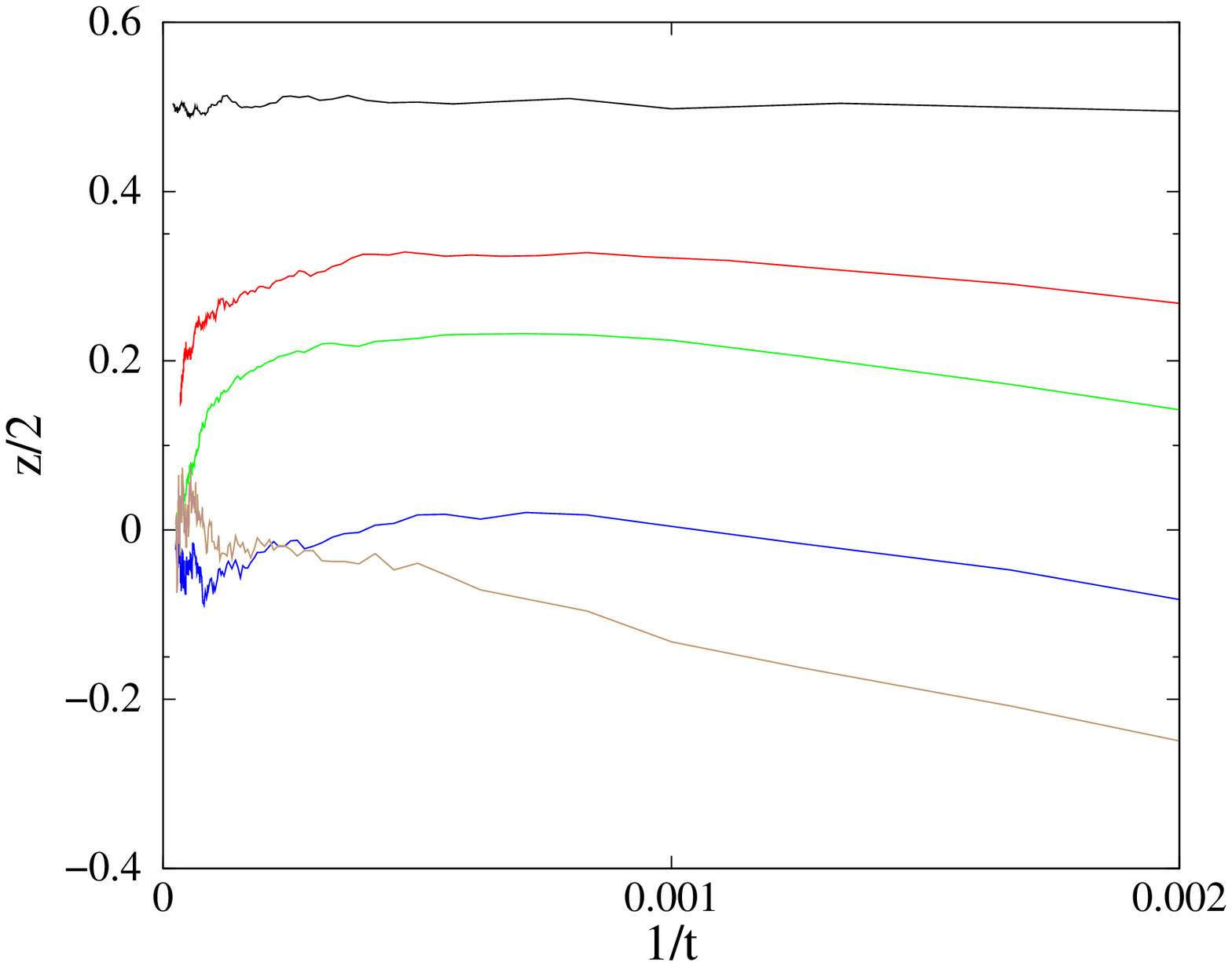}
\vspace{1mm}
\caption{Local exponent values of the '$+$'  cluster exponent  
$z/2 $ in the impure MDG case for $p_0=0, 0.15, 0.20, 0.35, 0.45$  
(from top to bottom).} 
\label{mdgplclz}
\end{figure}
\\
\\
\\
\begin{figure}[h]
\epsfxsize=80mm
\epsffile{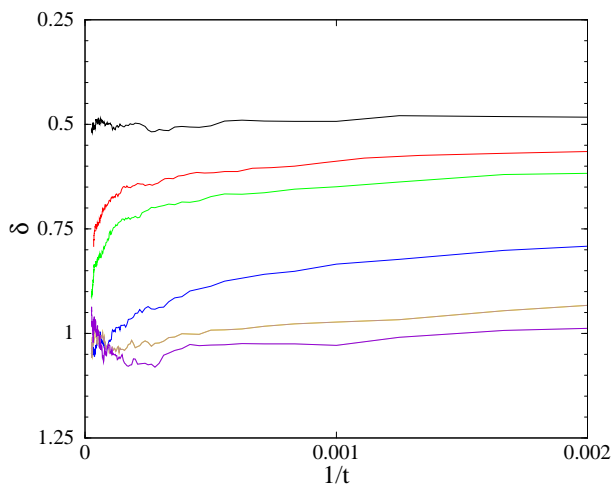}
\vspace{1mm}
\caption{Local exponent values of the '$+$' cluster exponent  
$\delta $ in the impure MDG case for $p_0=0, 0.15, 0.20, 0.35, 0.45$
(from top to bottom).} 
\label{mdgplcld}
\end{figure}

\section{Discussion and Conclusions}

In this paper the problem of the effect of quenched impurities
on one-dimensional non-equilibrium Glauber-Ising-type models has been
investigated numerically. Our previous  study of the same problem, 
without spin-anisotropy, led to the conclusion of a minimal of 
effect \cite{om05}. Here the emphasis has been on introducing 
spin-anisotropy, and to give insight into the the joint effect of 
kinetic and quenched spatial anisotropies. We have restricted our 
studies to the zero branching rate limit of the model (ARW-case). 
Depending on the ratio of anisotropies in the diffusion and 
annihilation channels (rates) a variety of behaviors appear when 
the quenched randomness is turned on, with the inference that 
some amount of kinetic anisotropy can significantly enhance 
the influence of quenched spatial anisotropies.

This behavior is different from that found for isotropic NEKIM 
\cite{om05}, where reaction and diffusion are completely blocked 
for very strong disorder. The effects of such blocking are investigated 
in the context of the $AB\to\emptyset$ model;
only the diffusive ones are found to be marginally relevant.

Summarizing the results for exponents $\eta$, $\delta$, and $z$
(Table \ref{clusres}), for '$+$' spins in the case $p= p_+ = 1/2$ (MDG+)
we have the $(0, 1/2, 1)$ spreading fixed point exponents of the
$XX\to X$ diffusive coagulation process, because due to Eq.(\ref{mdgrat}) 
the $XX\leftrightarrow X$ is unbalanced, favoring $XX\to X$ annihilation
of '$+$'-es and the proliferation of '$-$'-s.
In the $p_+ < p = 1/2$ kinetic disorder case (SAGI+) we found 
\cite{mo03} a novel kind of anisotropic fixed point $(1, 0, 2)$, which is the 
same what one obtains if quenched disorder is added to (MDG-). 
This can be understood by looking closer on the effect of these 
disorders on the spin species. Both kind of disorders enhance the
production of one species at the expense of the other. 
While in case of (SAGI+) the '$-$'-es are suppressed, for
the (diso MDG-) the same thing happens with the '$+$'-es (due to
Eq.(\ref{p0}) and the other spin domains grow linearly with $\eta=z/2=1$.

However due to Eq.(\ref{mdgrat}) the handicapped species end up in a
different way. In case of SAGI the '$-$'-s can survive 
resulting in $(0,0,0)$ as $t\to\infty$, slowing down the domain wall
decay (resulting in $\alpha=0.5$). 
In case of the diso-MDG the lonely '$+$'-s can die out 
completely, characterized by $\eta=-1$ mass and $\delta=1$ 
survival exponents, enabling the linear growth of kinks. 
Therefore $\alpha=1$ can be observed here.

\section{Acknowledgments}
Support from the Hungarian Research Fund OTKA (Grant No. T-046129) 
during this study is gratefully acknowledged. G. \'O. thanks the 
access to the NIIFI Cluster-GRID, LCG datagrid and the 
Supercomputer Center of Hungary.

\section {Appendix}

In this appendix the effect of quenched disorder in a further ARW type
model, the $AB\to\emptyset$ one, will be investigated. 
Unlike SAGI, this model is free of anisotropies.
In our previous paper \cite{om05} we investigated the inactive
phase of the cellular automaton version of the nonequilibrium 
kinetic Ising model (NEKIMCA) model with quenched disorder. In
that model  annihilating random walk of domain walls (kinks) 
$AA\to\emptyset$ is the dominant process. 
We have not found any effect of the weak disorder on the density decay
of kinks. If the disorder was so strong that diffusion could be blocked
at certain sites (the system is split into independent blocks of
finite sizes) a new fluctuating phase with finite steady state density
emerged as the consequence of the parity conservation of kinks.
Along and just below the blocking transition line the disorder caused 
logarithmic slow-down corrections to the $\rho \propto t^{-1/2}$ 
impure law of the ARW process \cite{Lee} and continuously changing exponents
($\alpha <0.5$) for strong disorder. Those results are in agreement with
analytic results for diffusion disorder \cite{SM98} and annihilation
disorder \cite{DM99} of the ARW.
\begin{figure}[ht]
\epsfxsize=80mm
\epsffile{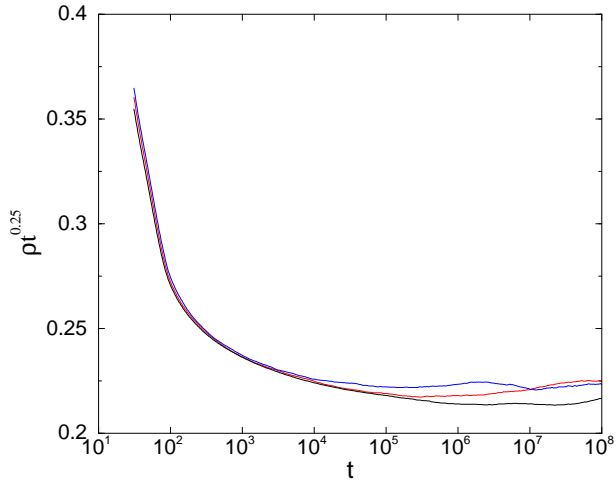}
\caption{Density decay of the annihilation disordered $AB\to\emptyset$
model in case of RI conditions \label{AB0qar} for
$\epsilon_A= 1, 0.98, 0.95$ (top to bottom)}
\end{figure}

The anisotropic reactions of the NEKIMA model \cite{mo02} result in a 
two component reaction dynamics of kinks destroying the PC class
critical behavior. Without kink production ($p_{ex}=0$) one can map
onto the $AB\to\emptyset$, however, since $A$ and $B$ species are
domain walls of the spins they are arranged alternately along the 1d
system. Consequently they can always meet  their nearest neighbors
(no hard-core reactions take place as in \cite{Mexproccikk})
and one observes the usual $\rho \propto t^{-1/2}$ decay instead
of the $\rho \propto t^{-1/4}$ law of the two component model \cite{LC95}.
\begin{figure}[ht]
\epsfxsize=80mm
\epsffile{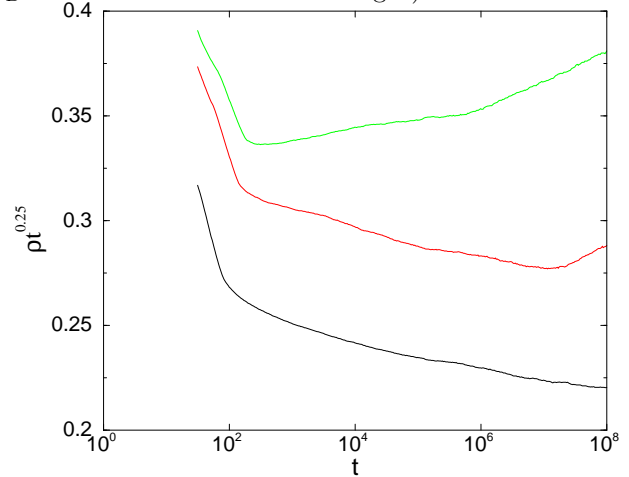}
\caption{Density decay of the diffusion disordered $AB\to\emptyset$
model in case of RI conditions \label{AB0rqd}. Different curves
correspond to $\epsilon_D = 1, 0.95, 0.5$ (top to bottom).}
\end{figure}

In this appendix we show results for the quenched disordered
$AB\to\emptyset$ model to complement results for the quenched
disorder NEKIMA model.  To our knowledge this has not been studied before.
We have investigated the effect of disorder in diffusion and annihilation
separately. A  random sequential simulation program has been run in 1d
systems of size $L=5\times 10^5$ with periodic boundary conditions 
and single particle occupation restriction. 
We have investigated the density decay for initial 
conditions with randomly (RI) and pair-wisely (PI) 
distributed $A$-s and $B$-s of full lattice occupancy. 

One elementary MC step is built up as follows. A particle and a direction is 
chosen randomly. If the nearest neighbor (nn) site in the selected 
direction was empty the particle is moved there with probability 
$1-\epsilon_D$ (where $\epsilon_D$ is the diffusion disorder
strength). If the nn site was occupied with a particle of a different type
we removed both particles with probability $1-\epsilon_A$.
The time ($t$) is incremented by $\Delta t = 1/n_p$
(where $n_p$ is the total number of particles) and the density decay 
was followed up to $t_{max}=10^8$ MCS.

In case of $\epsilon_D=0$ we did not see any effect of the disorder in 
the annihilation up to $\epsilon_A\le 1$ both for RI and PI conditions
(see Fig.\ref{AB0qar} for RI).
The density decay is completely insensitive to the quenched randomness
of the annihilation reaction rate.

On the other hand for disorder in the diffusion (hopping)
probabilities ($\epsilon_A=0$, $\epsilon_D \le 1$) one can see
logarithmic corrections to the scaling of the impure case for 
$\epsilon_D \to 1$ (see Fig. \ref{AB0rqd} for RI).
Following a sharp initial decay, when local correlations are built up
the evolution tends towards the $\rho\propto t^{-0.25}$ law, but as
in case of the ARW model logarithmic slow-down appears for very
late times and strong disorder (already without diffusion blocking).
It has been tested by simulating on different sizes ($L$) that the 
turns in the curves are not artifacts of finite system sizes.
Since the basic mechanism of annihilating random walk here 
$AB\to\emptyset$ is the same as for the $AA\to \emptyset$
model, we think that the relevancy of diffusion disorder
compared to the annihilation disorder can be observed
in the inactive phase of the NEKIMCA \cite{om05} as well.

\end{document}